\documentstyle[12pt]{article}
\begin{document}

Lietuvos fizikos \v zurnalas, 1996, {\bf 36,} Nr.4, 343-345

Lithuanian Physics Journal, 1996, {\bf 36}, No.4,

\vspace*{2.0\baselineskip}

\begin{center}
{\large {\bf Dynamical peculiarities of the nonlinear quasiclassical systems
}}\\[\baselineskip] B. Kaulakys\vspace*{0.5\baselineskip}\\ {\small
Institute of Theoretical Physics and Astronomy, A. Go\v stauto 12, 2600
Vilnius, Lithuania}\vspace*{1.5\baselineskip}\\[\baselineskip]
\parbox{5.5in}
{The quantum-classical correspondence for dynamics of the nonlinear
classically chaotic systems is analysed. The problem of quantum chaos
consists of two parts: the quasiclassical quantisation of the chaotic
systems and attempts to understand the classical chaos in terms
of quantum mechanics. The first question has been partially solved by
the Gutzwiller semiclassical trace formula for the eigenvalues of
chaotic systems, while the classical chaos may be derived from quantum
equations only introducing the decoherence process due to interaction
with system's environment or intermediate frequent measurement.
We may conclude that continuously observable quasiclassical system
evolves essentially classically-like. }
\end{center}

\vspace*{1.5\baselineskip}

Quasiclassical systems are systems with high quantum numbers and they may be
described quantum mechanically using the Bohr-Sommerfeld or WKB quantization
method. The quasiclassical (semiclassical) methods are built on the
classical trajectories. If a classical mechanical system can be represented
as multiperiodic, than it is equivalent to as many independent systems as
many degrees of freedom, or as many constants of the motion the system has.
Einstein, quoting Poincare, pointed out that such a separation was not
possible in the three-body problem and for, what he called, ''type B''
(chaotic) classical motion. However this statement of the Einstein was
completely ignored for many years.

Classical trajectories of the nonlinear systems are extremely complex, even
chaotic. Namely the classical dynamical chaos has prevented the broad
application of semiclassical ideas and techniques. After much effort Bohr,
Krammers, Born and Heisenberg failed in their attempt to quantize the helium
atom. They neither understood that classical helium atom is chaotic nor
understood dynamical chaos in general. Only in 1970 Gutzwiller derived his
semiclassical trace formula for the eigenvalues of chaotic systems (see, [1]
and references therein). Quasiclassical theory is an approximation of the
quantum mechanics and works very well in a great variety of systems and
large range of parameters, e.g. results to the very precise approach for the
matrix elements [2]. That is way the semiclassical theory, being really
worthwhile for the better understanding of many atomic, molecular and
mesoscopic processes has been named ''postmodern quantum mechanics''[3].
However it does not solve all problems of the quantum-classical
correspondence for the dynamics of the nonlinear systems.

In general, the quantum and classical realms are related by the
correspondence principle: physical characteristics of the highly exited
quantum systems with large quantum numbers are close to those of its
classical counterpart. However, recent research have suggested that the use
of the correspondence principle for strongly driven nonlinear systems is not
so straightforward: a quantum interference effect suppresses the classical
diffusion-like chaotic motion and results in the interference of macroscopic
systems' states (the Scr\"odinger cat paradox). The exponential instability
of classical motion characterized by the positive maximal Lyapunov exponent
$\Lambda >0$ destroys the deterministic image of the classical physics and
results in the unpredictability of the trajectory of the motion. [4]. The
largest Lyapunov exponent is defined as
$$
\Lambda =\lim \limits_{\left| t\right| \rightarrow \infty }\lim
\limits_{\varepsilon \left( 0\right) \rightarrow 0}\frac 1{\left| t\right|
}\ln \left| \frac{\varepsilon \left( t\right) }{\varepsilon \left( 0\right) }
\right| \eqno{(1)}
$$
where $\varepsilon \left( 0\right) $ is the variance in the phase space of
the initial conditions and $\varepsilon \left( t\right) $ is the tangent
vector of the trajectory at time moment $t.$ $\varepsilon \left( t\right) $
may be calculated from the linearised equations of motion. The exponential
instability implies a continuous spectrum of motion. On the other hand, the
classical phase space is continuous too.

The problem of the quantum chaos arise from attempts to understand the
classical chaos in terms of quantum mechanics. First at all, the energy and
frequency spectrum of any quantum system, which motion is bounded in phase
space, is always discrete, as well as the phase space. Accordingly, the
motion of such a system is regular. In addition, note to complications and
incompatibilities arising in attempts to define the quantum Lyapunov
exponent and KS (after Kolmogorov and Sinai) entropy [5]. Therefore, the
question: can (and if yes, how) quantum mechanics give chaos as a limiting
behaviour, is open until now. Here we only indicate to some possible
directions of search of the necessary and sufficient conditions for the
correspondence between quantum linear and classical nonlinear dynamics.

One possibility to overcome barriers posed by classical chaos between
quantum and classical dynamics is through the introduction of some
additional, usually nonlinear, terms to the quantum equations of motion.
Another approach is based on the coupling of the quantum system to
environment. As a result of interaction of the quasiclassical system with
the surrounding environment, the eigenstates of some observables
continuously decohere and can behave like classical states [6].

To facilitate the comparison between quantum and classical dynamics it is
convenient to employ the Wigner representation for the density matrix $\rho
\left( x,p,t\right) $. The Wigner function of the system evolves according
to equation
$$
\frac{\partial \rho }{\partial t}=\left\{ H,\rho \right\} _M\equiv \left\{
H,\rho \right\} +\sum\limits_{n\geq 1}\frac{\hbar ^{2n}\left( -1\right) ^n}{
2^{2n}\left( 2n+1\right) !}\frac{\partial ^{2n+1}V}{\partial x^{2n+1}}\frac{
\partial ^{2n+1}\rho }{\partial p^{2n+1}},\eqno{(2)}
$$
where by $\left\{ ...\right\} _M$ and $\left\{ ...\right\} $ are denoted the
Moyal and the Poisson brackets respectively, while the Hamiltonian of the
system is of the form $H=p^2/2m+V\left( x,t\right) $. The terms in eq. (2)
containing Planck's constant give the quantum corrections to the classical
dynamics generating by the Poisson brackets. These terms contain higher
derivatives. In the region of regular dynamics one can neglect the quantum
corrections for very long time if the characteristic actions of the system
are large [7,8]. In case of chaotic classical motion the exponential
instabilities lead to the development of the fine structure of the
distribution function and exponential grow of its derivatives. As a result,
the quantum corrections in eq. (2) become significant after relatively short
time even for macroscopic bodies, e.g. Hyperion, the chaotically turning
moon of Saturn [8-10]. However, Hyperion evolves perfectly classically,
because interaction with environment causes the loss of quantum coherence
(superposition of the macroscopic states) and results in momentum diffusion.
Thus the physical mechanism which reduces the quantum dynamics of a
mesoscopic or macroscopic particle to the classical dynamics is the
decoherence process due to the interaction with the environment. It is
impossible to isolate macroscopic systems from their environments for a time
comparable to their dynamical time scale, furthermore, even the internal
degrees of freedom may play the role of the environment [7]. Under some
sufficiently general assumptions the environment's influence reduces to the
equation of quantum evolution with some diffusion term responsible for the
decoherence process, i.e. decay of the off-diagonal terms of the density
matrix.

On the other hand, frequent measurements of the quasiclassical systems
results to the classical diffusion-like motion too [11,12]. Repeated
frequent measurement of suppressed systems results in the delocalization.
Time evolution of the observable chaotic systems becomes close to the
classical frequently broken diffusion-like process described by rate
equations for the probabilities rather than for amplitudes [11,12]. Note
that continuous quantum monitoring of the system may be represented as
restrictions of the quantum trajectories in the path integral representation
of quantum mechanics [13]. Therefore, the quantum evolution of frequently
(continuously) observable chaotic system interacting with the detector or
environment is more classical-like than evolution of the idealised isolated
system.

However, as it has been noted by Kolovsky [7], it seems to be impossible to
derive the Liouvile equation from quantum mechanics by any way and, maybe,
''the Liouvile equation (as well as the Hamiltonian equations) is a sort of
approximation''. The extremely small additional diffusion term which takes
into account the influence of the internal and external quantum motion
prohibits development of the distribution function's ''fine structure'' and
removes barriers posed by classical chaos for the correspondence principle.
Some necessary conditions which can lead to the occurrence of decoherence in
a system interacting with an apparatus or environment are analysed in a
recent paper [14].

In conclusion, the quantum-classical correspondence problem caused of the
chaotic dynamics is closely related with the old problem of measurement in
quantum mechanics. Even the simplest detector follows irreversible dynamics
due to the coupling to the multitude of vacuum modes which results in the
decay of the phases of the quantum amplitudes or off-diagonal matrix
elements of the density matrix--what we need to obtain the classical
equations of motion.

\vspace{0.2cm} {\bf Acknowledgment.} The research described in this
publication was made possible in part by Grant No. LHV100 from the Joint
Fund Program of Lithuanian Government and International Science Foundation.

\vspace{1.5cm}{\large References} \vskip\baselineskip

\begin{enumerate}
\item  M. C. Gutzwiller, Chaos in classical and quantum mechanics
(Springer-Verlag, 1990).

\item  B. Kaulakys, ''Consistent analytical approach for the quasiclassical
radial dipole matrix elements'', J. Phys. B: At. Mol. Opt. Phys., V. 28, No.
23, p. 4963-71 (1995).

\item  E. J. Heller and S. Tomsovic, ''Postmodern quantum mechanics'', Phys.
Today, V. 46, No. 7, p. 38-46 (1993).

\item  G. Casati and B. Chirikov, ''The legacy of chaos in quantum
mechanics'', in Quantum chaos: between order and disorder, Ed. G. Casati and
B. V. Chirikov (Cambridge University, 1994).

\item  W. Slomczynski and K. \v Zyczkowski, ''Quantum chaos: an entropy
approach'', J. Math. Phys., V. 35, No. 11, p. 5674-5700 (1994).

\item  W. H. Zurek, ''Decoherence and the transition from quantum to
classical'', Phys. Today, V. 44, No. 10, p. 36-44 (1991).

\item  A. R. Kolovsky, ''A remark on the problem of quantum-classical
correspondence in case of chaotic dynamics'', Europhys. Lett., V. 27, No. 2,
p. 79-84 (1994).

\item  W. H. Zurek and J. P. Paz, ''Decoherence, chaos, and second law'',
Phys. Rev. Lett., V. 72, No. 16, p. 2508-11 (1994).

\item  G. Casati and B. V. Chirikov, ''Comment on 'Decoherence, chaos, and
second law''', Phys. Rev. Lett., V. 75, No. 2, p. 350 (1995).

\item  W. H. Zurek and J. P. Paz, ''Reply to Comment...'', ibid, p. 351
(1995).

\item  B. Kaulakys, ''On the quantum evolution of chaotic systems affected
by repeated frequent measurement'', In: Quantum Communications and
Measurement, Ed. V. P .Belavkin et al, (Plenum Press, New York, 1995)
p.193-197.

\item  V. Gontis and B. Kaulakys, ''Quantum dynamics of simple and complex
systems affected by repeated measurement'', in Proc. Intern. Conf. on
Nonlinear Dynamics, Chaotic and Complex Systems (NDCCS'95), 7-12 Nov. 1995,
Zakopane, Poland (to be published).

\item  M. B. Mensky, R. Onofrio and C. Presilla, ''Continuous quantum
monitoring of position of nonlinear oscillators'', Phys. Lett.A, V. 161, No.
3, p. 236-40 (1991).

\item  M. Dugi\v c, ''On the occurrence of decoherence in nonrelativistic
quantum mechanics'', Phys. Scripta, V. 53, No. 1, p. 9-17 (1996).
\end{enumerate}

\end{document}